\documentclass[a4paper, 10pt]{article}

\usepackage{cite}
\usepackage{amsmath,amssymb,amsfonts}
\usepackage{algorithmic}
\usepackage{graphicx}
\usepackage{textcomp}
\usepackage[para]{footmisc}
\usepackage{xcolor}

\def\BibTeX{{\rm B\kern-.05em{\sc i\kern-.025em b}\kern-.08em
    T\kern-.1667em\lower.7ex\hbox{E}\kern-.125emX}}

\newcommand{\abele}{{\scshape abele}}
\newcommand{\lime}{{\scshape lime}}
\newcommand{\lore}{{\scshape lore}}

\newcommand{\shap}{{\scshape shap}}

\usepackage[pscoord]{eso-pic}

\begin{document}

\title{Exemplars and Counterexemplars Explanations for Image Classifiers, Targeting  Skin Lesion Labeling}





\author{
Carlo Metta
    \thanks{ISTI-CNR, Pisa, Italy}
    \and
    Riccardo Guidotti
    \thanks{University of Pisa, Italy}
    \and
     Yuan Yin
    \thanks{Sorbonne Universite, Paris, France}
    \and
    Patrick Gallinari
    \footnotemark[3]
    \and
    Salvatore Rinzivillo
    \footnotemark[1]
}

\maketitle

\begin{abstract}
Explainable AI consists in developing mechanisms allowing for an interaction between decision systems and humans by making the decisions of the formers understandable. This is particularly important in sensitive contexts like in the medical domain. We propose a use case study, for skin lesion diagnosis, illustrating how it is possible to provide the practitioner with explanations on the decisions of a state of the art deep neural network classifier trained to characterize skin lesions from examples. Our framework consists of a trained classifier onto which an explanation module operates. The latter is able to offer the practitioner exemplars and counterexemplars for the classification diagnosis thus allowing the physician to interact with the automatic diagnosis system. The exemplars are generated via an adversarial autoencoder.  We illustrate the behavior of the system on representative examples.  
\end{abstract}

Image classification, Explainable AI, Machine Learning, Skin Lesion Image Classification, Adversarial Autoencoders

\section{Introduction}
In the last years, AI based decision support systems have gained a huge impact in different domains, in many cases providing high accuracy predictions, classification, and recommendation. However, their adoption in high-stake scenario that involves decision on humans has raised several ethical concerns about the fairness, bias, transparency and dependable decisions taken on the basis of AI suggestions~\cite{pedreschi2019meaningful}. These concerns are even more relevant in mission-critical domains, like in healthcare. Thus, it is necessary to develop AI systems that are able to assist the doctors to take informative decisions, complementing their own knowledge with the information and suggestion yielded by the AI system.

Our proposal starts from the design and development of a classification model for the \textit{ISIC 2019 Challenge}. The objective of the challenge is to invite research to develop automated systems to provide accurate classification of skin cancers from dermoscopic images. Our system consists of two main modules: \textit{a)} a CNN model to classify each input image as a class  among eight possibilities; \textit{b)} an explainer based on exemplars and counter exemplars synthesis that exploit and adversarial autoencoder (AAE) to produce the images for the explanation. In this paper we introduce the models and their structures. In particular, we have used two separate learning processes: \textit{a)} a learning phase for the CNN, starting from the ResNet50 architecture; and \textit{b)} a training phase for an Adversarial AutoEncorder (AAE). Since the image produced by the AAE are intended to be used as exemplars, it is crucial to have a wide catalog of neighborhood instances. For this reason, we developed a progressive growing AAE to maximize the diversification of the generated images.
The two training processes are intentionally kept separated since two different loss functions are used (the first for accuracy, the second for discrimination). We also want to demonstrate that the explainer may be effective even if the original training set was not available.
In the paper we show that accurately designing the AAE is crucial for obtaining an explanation based on realistic exemplars and counter-exemplars, especially in a specialistic domain as healthcare.

The rest of the paper is organized as follows.
In Section~\ref{sec:related} we list recent contributions in the field of explainability for image classification.
Section~\ref{sec:background} recalls basic notions related with the approach exploited in this paper.
In Section~\ref{sec:casestudy} we illustrate in detail the settings of the case study addressed, while Section~\ref{sec:explanations} shows the results.
Finally, Section~\ref{sec:conclusion} summarizes the contribution and proposes future research directions.

\section{Related Work}
\label{sec:related}
Image classification can be widely applied in health for various purposes ranging from heart disease diagnosis to skin cancer detection~\cite{iqbal2021automated,li2014medical,antonie2021Application}.
In order to have high performance in these tasks, AI systems relying on machine learning models are more and more used.
Unfortunately, these models are typically \textit{black boxes} handing the rationale of their behavior.
For this reason, research on \textit{black box explanation} has recently received much attention~\cite{adadi2018peeking,miller2019explanation,guidotti2019survey}.
This interest is driven by the idea of adopting into AI systems explanation methods such that high performance and interpretability can coexist.
Explainability is practically useful in social sensitive context like the medical one~\cite{panigutti2020doctor}.

In image classification, typical explanations are \textit{saliency maps}, i.e., images that show each pixel's positive (or negative) contribution.
At a high level, explanation methods can be categorized as \textit{model-specific or model-agnostic}, depending on whether the explanation method exploits knowledge of the internal structure of the black box or not; \textit{global or local}, depending on whether the explanation is provided for the black box as a whole or for any specific instance.
The explainer \abele{} adopted in this paper is a \textit{local} \textit{model-agnostic} method.

With respect to the literature, \lime{}~\cite{ribeiro2016should} and \shap{}~\cite{lundberg2017unified} are two of the most well known data agnostic local explanation methods.
\lime{} randomly generates a local neighborhood ``around'' the instance to explain, labels them using the black box under analysis and returns an explanation using as surrogate model a linear regressor.
On the other hand, \shap{} adopts game theory and exploits the Shapley values of a conditional expectation function of the black box, providing for each feature the unique additive importance.
Besides being model-agnostic, \lime{} and \shap{} are also theoretically not tied to a specific type of data.
Indeed, they can be applied to explain image classifiers and return explanation in the form of saliency maps by turning the features importance into pixels importance.
They achieve this objective by using for the process ``superpixels'', i.e., areas of the image under analysis with similar colors.
Unfortunately, the usage of superpixels requires a segmentation procedure that affects the explanation.
Moreover, the neighborhoods considered when investigating the black box behavior are no longer plausible instances but simply the image under analysis with some pixels ``obscured''~\cite{guidotti2019investigating}.
The fact that the explanation procedure relies on not plausible images is generally unpleasant in medical applications.
\abele{} overcomes these issues relying on a realistic procedure for generating images similar to the one under analysis and does not require any a priori segmentation~\cite{guidotti2019black}.

Other explanation methods widely used to build saliency maps are model-specific approaches such as IntGrad~\cite{sundararajan2017axiomatic}, GradInput~\cite{shrikumar2016not}, and $\varepsilon$-LRP~\cite{bach2015pixel}.
In brief, to retrieve the saliency map, they redistribute the prediction backwards using local rules until it assigns a relevance score to each pixel value.
These approaches are designed for deep neural networks and cannot be employed for explaining image classifiers for which the type of the model is unknown.
On the other hand, being model-agnostic, \abele{} overcomes this limitation and also allows playing with the components of the explanation while these model-specific approaches are more limited.

\begin{figure}[t]
    \centering
    \includegraphics[width=\columnwidth]{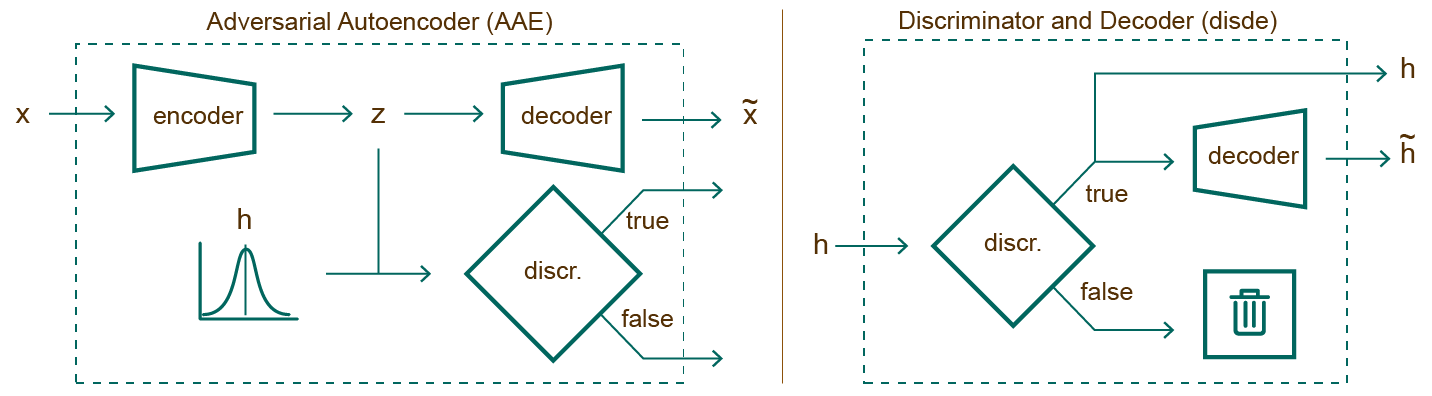}
    \caption{\textit{Left}: AAE architecture. \textit{Right}: Discriminator and Decoder module.}
    \label{fig:AAE}
\end{figure}

\section{Setting The Stage}
\label{sec:background}

\subsection{Adversarial Autoencoders}
\label{sec:AAE}
An important issue arising in the use of synthetic instances when developing black box explanations is the question of maintaining the identity of the distribution of the examples that are generated with the prior distribution of the original examples. 
In \cite{guidotti2019black} this issue is approached by using an Adversarial Autoencoder (AAE)~\cite{makhzani2015adversarial}, which combines a Generative Adversarial Network (GAN)~\cite{goodfellow2014generative} with the autoencoder representation learning algorithm.

AAEs are probabilistic autoencoders that aim at generating new random items that are highly similar to the training data. 
They are regularized by matching the aggregated posterior distribution of the latent representation of the input data to an arbitrary prior distribution. 
The AAE architecture (Figure~\ref{fig:AAE}-left) includes an $\mathit{encoder}: \mathbb{R}^n {\rightarrow} \mathbb{R}^k$, a $\mathit{decoder}: \mathbb{R}^k {\rightarrow} \mathbb{R}^n$ and a $\mathit{discriminator}: \mathbb{R}^k {\rightarrow} [0, 1]$ where $n$ is the number of pixels in an image and $k$ is the number of latent features.
Let $x$ be an instance of the training data, we name $z$ the corresponding latent data representation obtained by the $\mathit{encoder}$.
We can describe the AAE with the following distributions~\cite{makhzani2015adversarial}: the prior distribution $p(z)$ to be imposed on $z$, the data distribution $p_d(x)$, the model distribution $p(x)$, and the encoding and decoding distributions $q(z|x)$ and $p(x|z)$, respectively.
The encoding function $q(z|x)$ defines an aggregated posterior distribution of $q(z)$ on the latent feature space: $q(z) {=} \int_x q(z|x) p_d(x) dx$.
The AAE guarantees that the aggregated posterior distribution $q(z)$ matches the prior distribution $p(z)$, through the latent instances and by minimizing the reconstruction error. 
The AAE generator corresponds to the encoder $q(z|x)$ and ensures that the aggregated posterior distribution can confuse the $\mathit{discriminator}$ in deciding if the latent instance $q(z)$ comes \mbox{from the true 
distribution $p(z)$.}

The AAE learning involves two phases: the \textit{reconstruction} aimed at training the $\mathit{encoder}$ and $\mathit{decoder}$ to minimize the reconstruction loss; the \textit{regularization} aimed at training the $\mathit{discriminator}$ using training data and encoded values.
After the learning, the decoder defines a generative model mapping the prior distribution $p(z)$ to distribution 
$p_d(x)$.

\subsection{ABELE}
\abele{} (Adversarial Black box Explainer generating Latent Exemplars) is a local model agnostic explainer for image classifiers~\cite{guidotti2019black}.
Given an image $x$ to explain and a black box $b$, the explanation provided by \abele{} is composed of \textit{(i)} a set of \textit{exemplars} and \textit{counter-exemplars}, \textit{(ii)} a \textit{saliency map}.
Exemplars and counter-exemplars show instances classified with the same outcome as $x$, and with an outcome other than $x$, respectively.
They can be visually analyzed to understand the reasons for the decision.
The saliency map highlights the areas of $x$ that contribute to its classification and areas that push it into another class.
The explanation process is as follows.
First, \abele{} generates a neighborhood in the latent feature space exploiting an Adversarial Autoencoder (AAE)~\cite{makhzani2015adversarial}.
Then, it learns a decision tree on that latent neighborhood providing local decision and counterfactual rules~\cite{guidotti2019factual}.
Finally, \abele{} selects and decodes exemplars and counter-exemplars satisfying these rules and extracts from them a saliency map.

\begin{figure}[t]
    \centering
    \includegraphics[width=\columnwidth]{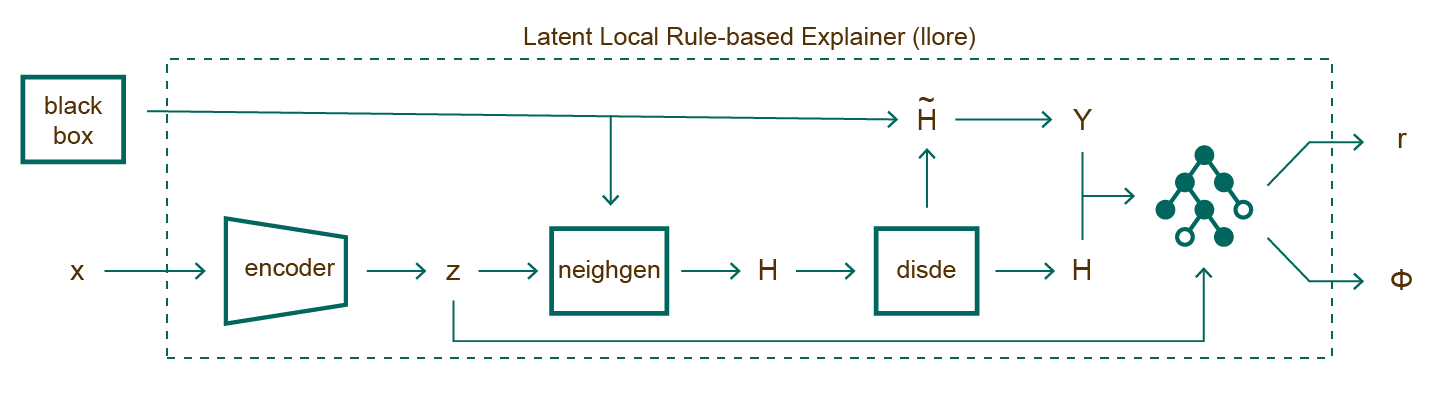}
    \caption{Latent Local Rules Extractor ($\mathit{llore}$) module.}
    \label{fig:learning}
\end{figure}

\subsubsection{Encoding} 
The image $x {\in} \mathbb{R}^n$ to be explained is passed as input to the AAE where the $\mathit{encoder}$ returns the latent representation $z \in \mathbb{R}^k$ using $k$ latent features with ${k \ll n}$. 

\subsubsection{Neighborhood Generation} 
\abele{} generates a set $H$ of $N$ instances in the latent feature space, with characteristics close to those of $z$. 
Since the goal is to learn a predictor on $H$ able to simulate the local behavior of $b$, the neighborhood includes instances with both decisions, i.e., $H = H_{=} \cup H_{\neq}$ where instances $h \in H_=$ are such that $b(\widetilde{h}) = b(x)$, and $h \in H_{\neq}$ are such that $b(\widetilde{h}) \neq b(x)$. 
We name $\widetilde{h} \in \mathbb{R}^n$ the decoded version of an instance $h \in \mathbb{R}^k$ in the latent feature space.
The neighborhood generation of $H$ ($\mathit{neighgen}$ module in Fig.~\ref{fig:learning}) may be accomplished using different strategies ranging from pure random strategy using a given distribution to a genetic approach maximizing a fitness function~\cite{guidotti2019factual}. 
In our experiments we adopt the last strategy.
After the generation process, for any instance $h \in H$, \abele{} exploits the $\mathit{disde}$ module (Fig.~\ref{fig:AAE}-right) for both checking the validity of $h$ by querying the $\mathit{discriminator}$ and decoding it into $\widetilde{h}$.
Then, it queries the black box $b$ with $\widetilde{h}$ to get the class $y$, i.e.,~$b(\widetilde{h})=y$.

\subsubsection{Local Classifier Rule Extraction} 
Given the local neighborhood $H$, \abele{} builds a decision tree classifier $c$ trained on $H$ labeled with $b(\widetilde{H})$. 
The surrogate tree is intended to locally mimic the behavior of $b$ in the neighborhood $H$. 
It extracts the decision rule $r$ and counter-factual rules $\Phi$ enabling the generation of \textit{exemplars} and \textit{counter-exemplars}.
Fig.~\ref{fig:learning} shows the process that, starting from the image to be explained, leads to the decision tree learning, and to the extraction of the decision and counter-factual rules. 
We name this module \textsc{llore}, as a variant of \lore{}~\cite{guidotti2019factual}. 

\begin{figure}[t]
    \centering
    \includegraphics[width=\columnwidth]{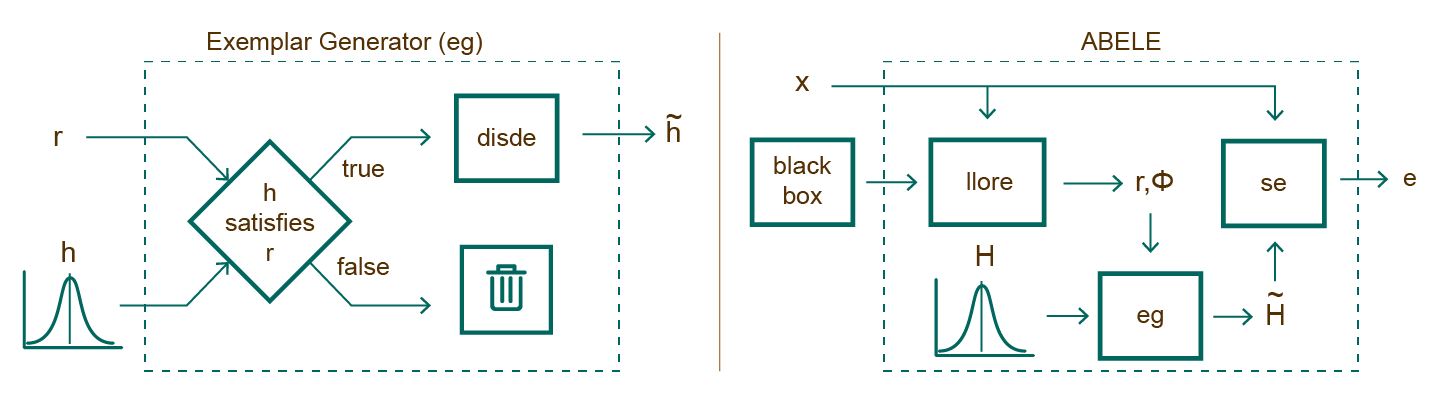}
    \caption{\textit{Left}: Exemplar Generator ($\mathit{eg}$) module. \textit{Right}: \abele{} architecture.}
    \label{fig:abele}
\end{figure}

\subsubsection{Explanation Extraction} 
Often, e.g.~in medical or managerial decision making, people explain their decisions by pointing to exemplars with the same (or different) decision outcome.
We follow this approach and we model the explanation of an image $x$ returned by \abele{} as a triple ${e=\langle \widetilde{H}_e, \widetilde{H}_c, s \rangle}$ composed by \textit{exemplars} $\widetilde{H}_e$, \textit{counter-exemplars} $\widetilde{H}_c$ and a \textit{saliency map} $s$.
Exemplars and counter-exemplars are images representing instances similar to $x$, leading to an outcome equal to or different from $b(x)$. 
Exemplars and counter-exemplars are generated by \abele{} exploiting the $\mathit{eg}$ module (Fig.~\ref{fig:abele}-left). 
It first generates a set of latent instances $H$ satisfying the decision rule $r$ (or a set of counter-factual rules $\Phi$), as shown in Fig.~\ref{fig:learning}.
Then, it validates and decodes them into exemplars $\widetilde{H}_e$ (or counter-exemplars $\widetilde{H}_c$) using the $\mathit{disde}$ module.
The saliency map $s$ highlights areas of $x$ that contribute to its outcome and areas that push it into another class.
The map is obtained by the saliency extractor $\mathit{se}$ module (Fig.~\ref{fig:abele}-right) that first computes the pixel-to-pixel-difference between $x$ and each exemplar in the set  $\widetilde{H}_e$, and then, it assigns to each pixel of the saliency map $s$ the median value of all differences calculated for that pixel.
Thus, formally for each pixel $i$ of the saliency map $s$ we have: 
$s[i] = \mathit{median}_{\forall \widetilde{h}_e \in \widetilde{H}_e}(x[i] - \widetilde{h}_e[i]).$

\section{Case Study}
\label{sec:casestudy}



We present here a description of all the components involved in the training process of the black box model used to classify instances of the ISIC dataset in eight different classes, and of the AAE used by \abele{} trained on the same dataset.

\subsection{ISIC Dataset, Prerocessing and Classifier}
Skin Lesion Analysis Towards Melanoma Detection is a challenge proposed by the International Skin Imaging Collaboration (ISIC), an international effort to improve melanoma diagnosis, sponsored by the International Society for Digital Imaging of the Skin (ISDIS). ISIC has developed an international repository of dermoscopic images, for both the purposes of clinical training, and for supporting technical research toward automated algorithmic analysis. The goal for ISIC 2019 is to classify dermoscopic images among nine different diagnostic categories: MEL (Melanoma), NV (Melanocytic nevus), BCC (Basal cell carcinoma), AK (Actinic keratosis), BKL (Benign keratosis), DF (Dermatofibroma), VASC (Vascular lesion), SCC (Squamous cell carcinoma), UNK (None of the others / out-of-distribution).

\begin{figure}[t]
    \centering
    \includegraphics[width=\columnwidth]{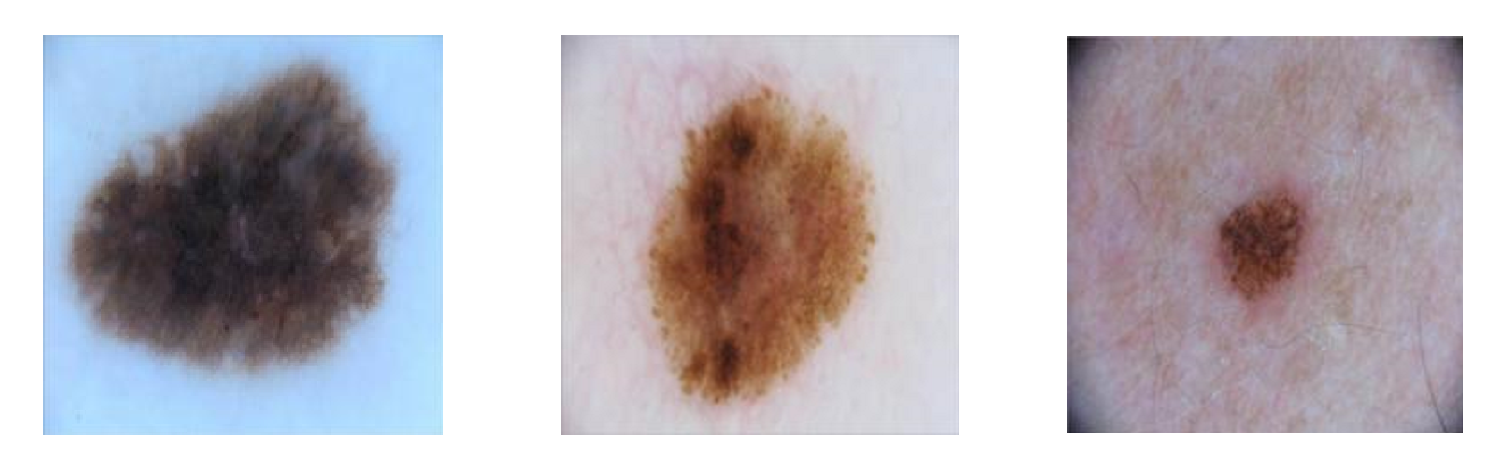}
    \caption{Melanocytic nevus preprocessed samples}
    \label{fig:datasamples}
\end{figure}

\subsubsection{Dataset}
The dataset is composed of
a \textit{training set} of 25,331 JPEG images of skin lesions and their category (labels);
a \textit{test set} of 8,238 unlabelled JPEG images of skin lesions.
In our experiments we relied on the training set for which ground truth is directly available. 
In particular, from the training set, we used $80\%$ for training and $20\%$ for validation. 
For the evaluation, we use the same evaluation protocols as the submission system. 

\subsubsection{Data Preprocessing}
Since the images have different resolutions. We preprocess them as follows:
\begin{itemize}
    \item For the training process, the images are randomly rescaled, rotated and cropped to generate the input to the network. Note that such preprocessing does not deform the lesions in the image. Resolution of the preprocessed images is 224×224.
    \item For the validation and test process, each image is firstly rescaled to 256×256 according to the shorter edge, then cropped at the center into a 224×224 image.
\end{itemize}
Since the UNK category is a reject option and is not available in the training we focused on the 8 other categories.

\subsubsection{Image Classifier Architecture}
For the classification, we used a classical ResNet, pretrained on Imagenet, i.e., a black box classifier.
In particular, we replaced the last classification layer by the new one adapted to the number of classes in the diagnosis. 
We trained this layer and fine tuned the rest of the network on ISIC dataset.
The $50$-layer ResNet architecture is the following: The network is composed of 18 modules sequentially combined together, including one \texttt{conv1} module (7×7, 64 filters, stride 2), three \texttt{conv2} modules (1×1, 64 filters, stride 1; 3×3, 64 filters, stride 1; 1×1, 256 filters, stride 1), four \texttt{conv3} (1×1, 128 filters, stride 2; 3×3, 128 filters, stride 1; 1×1, 512 filters, stride 1), six \texttt{conv4} (1×1, 256 filters, stride 2; 3×3, 256 filters, stride 1; 1×1, 1024 filters, stride 1), three \texttt{conv5} (1×1, 512 filters, stride 2; 3×3, 512 filters, stride 1; 1×1, 2048 filters, stride 1) and one last fully connected module \texttt{fc} (average pooling, 9-output fully connected layer, sigmoid activation). The first module \texttt{conv1} is composed of one single convolution layer. For \texttt{conv2} to \texttt{conv5}, each module is a residual block including three convolutional layers in the residual branch. The output of such block is the sum of the input and the output of the convolutional layers. The module \texttt{fc} is the newly trained prediction layer. The spatial size is reduced only at the first layer of the modules \texttt{conv3}, \texttt{conv4} and \texttt{conv5}.
We adopt a binary cross entropy loss for each class, so that the problem is considered as 8 individual one-vs-rest binary classification problems.

\subsubsection{Evaluation Criteria}
The official evaluation criterion for the challenge is the Normalized (or balanced) multi-class accuracy. 
It is defined as the average of recall obtained in each class. The best value is 1 and the worst is 0. This metric makes all the classes equally important.
In terms of balanced multi-class accuracy (mean value of categorical recalls), the trained model achieved 0.838 on the validation set.
We report also the performance on the official test set.
The score is 0.488, which is potentially impacted by the presence of out-of-distribution samples (UNK) since the model was not tuned for rejection.

\subsection{Customization of ABELE}
We describe here the customization of \abele{} we carried on in order to make it usable for the complex image classification task addressed by the ISIC classifier previously described.

\subsubsection{Issues with Generative Models}
Generative Adversarial models are generally not easy to train as they are usually affected by a number of common failures. 
These problems vary from a diversified spectrum of failures in convergence to the famous Mode Collapse \cite{modecollapse}, the tendency by the generator network to produce a small variety of output types.
Such problems mainly arise from the competing scheme generator and discriminator are trained on: they are trained simultaneously in a zero-sum game, thus the goal is to find an equilibrium between the two competing objectives. Since every time the parameters of one of the models are updated the nature of the optimization problem is changed, this has the effect of creating a dynamic system that easily fails to converge.

Furthermore, even the concept of convergence is not as clear as in other context. As the generator improves during the training, the discriminator performance gets worse because it cannot easily tell the difference between real and fake outputs. At the limit where the generator succeeds perfectly, the discriminator flips a coin to make its prediction. Hence, the discriminator feedback gets less meaningful over time. If the model continues training past the point when the discriminator is giving random feedback, then the generator starts to train on low quality feedback, and its own performance may collapse.
For a Generative Adversarial model, convergence is often an unstable transitory region rather than a stable state.
\\
In addition to the previous problems, we often face the further complication to deal with real world datasets that are far from ideal: fragmentation, imbalance, lack of uniform digitization, shortage of data are primary challenges of big data analytics for healthcare. All of them impede efficiency and accuracy of machine learning model trained with these data, especially in the case of 
inherently fragile generative models.

Training an AAE in a standard fashion to reproduce samples from ISIC dataset without taking special care of all issues mentioned above resulted in extremely poor performance, mostly due to a persistent collapse mode. In order to overcome such generative failure and dataset limitations, we implemented a collection of cutting edge techniques that altogether is capable of addressing all the issues we mentioned and successfully training an AAE with adequate performance.

\subsubsection{Overcoming Mode Collapse}
One usually wants a generative model to produce a large variety of outputs, for example, a different image for every input of our skin lesion generator. However, since the generator is always trying to produce the one output that seems most plausible to the discriminator, it may happen that the generator learns to produce just a single output over and over again, or a small set of outputs.
In such case, the discriminator best strategy is to learn to always reject that output. But if the current discriminator gets stuck in a poor local minimum it is not going to find the best strategy, hence it would be too easy for the generator to find the most credible output for the current discriminator. 

As a consequence, at each iteration of generator an over-optimization for the current discriminator takes place, and the discriminator never learns to escape out of the trap. Thus, each generation of generator rotates through a small set of output types, possibly a single one. In adversarial generative models, this form of failure is called mode collapse.

In recent years many techniques have been proposed to overcome this ubiquitous form of failure. Ad hoc tricks like Mini Batch Discrimination \cite{minibatch} or Wasserstein Loss \cite{wasserstein} have been proved empirically to alleviates mode collapse, while Unrolled GANs \cite{unrolled} or Conditional AAE \cite{conditional} intervene directly on the internal structure of the training scheme to discouraging over-optimization by the current generator.

The reasons behind mode collapse and other forms of failure remain still unclear and not fully proven. In the health domain, such singularities around the train process appear to be even more frequent. Possible reasons can be the lack of a substantial number of training data, the necessity to deal with high resolution images, fragmented and highly unbalanced datasets (malignant cancers are usually a small proportion of the entire batch). Furthermore, the need of a substantial number of latent features makes the parameter space extremely irregular, non-convex and in need of powerful regularization. In order to overcome failure modes and train an AAE successfully, we used the following collection of techniques.

\subsubsection{Progressive Growing Adversarial Autoencoder}
Progressive Growing GANs \cite{progressive} have been introduced as an extension to the GANs training process. It helps to achieve a more stable training of generative models for high resolution images. The main idea is to start with a very low resolution image and step by step adding block of layers that simultaneously increase the output size of the generator model and the input size of the discriminator model until the desired size is achieved.

In a general GAN scheme, the discriminator is linked to the generator model output. However, in an AAE the discriminator takes as input the encoded latent space instead of the full reconstructed image. In order to achieve all benefits of a progressive growing model, we need a novel structure to take care of this different dynamics. We then propose a Progressive Growing Adversarial Autoencoder (PGAAE): starting with a single block of convolutional layers for each of the two generating networks (encoder and decoder) we are able to reconstruct low resolution images (7x7 pixels), then step by step we increase the number of blocks until the network is powerful enough to manage images of the desired size (224x224 pixels). The latent space dimension is kept fixed, consequently the discriminator takes as input tensors always the same size. Although one could fix also the discriminator network, we found helpful to progressively increasing also the width of this network so that the discriminator can deal each step with a more structured information. On the contrary, we observe that increasing the depth of the discriminator increases the instability of the training, causing a wide variety of failures ranging from poor performance to catastrophic forgetting \cite{modecollapse}.

The main idea behind such construction relies on the instability caused to the training process by heavy structured high dimensional data.
Generating high-resolution images is challenging for generative models, as the generator must learn how to output both high dimensional structure and fine details at the same time. The high resolution makes any discrepancy in the fine detail of generated images easy to mark for the discriminator, and the training process degenerates. Large images also require significantly more memory, consequently, the batch size used to update model weights each iteration is reduced to ensure that the large images fit into memory. This introduces further instability into the process.

The incremental addition of the layers allows the models to first learn large scale structure and progressively shift the attention to finer detail. At each step, all previous layers remain trainable throughout the training process. From a different point of view, one can think of each block of layers as the initialization of the same common network structure of the subsequent step of the progressive scheme. Such progressive initialization is by all means a powerful form of regularization carried out through both the encoder and decoder networks. A heavy quantity of regularization is indeed needed to smooth the parameter space and, in turn, reducing failure modes like mode collapse.

\begin{figure}[t]
    \centering
    \includegraphics[width=\columnwidth]{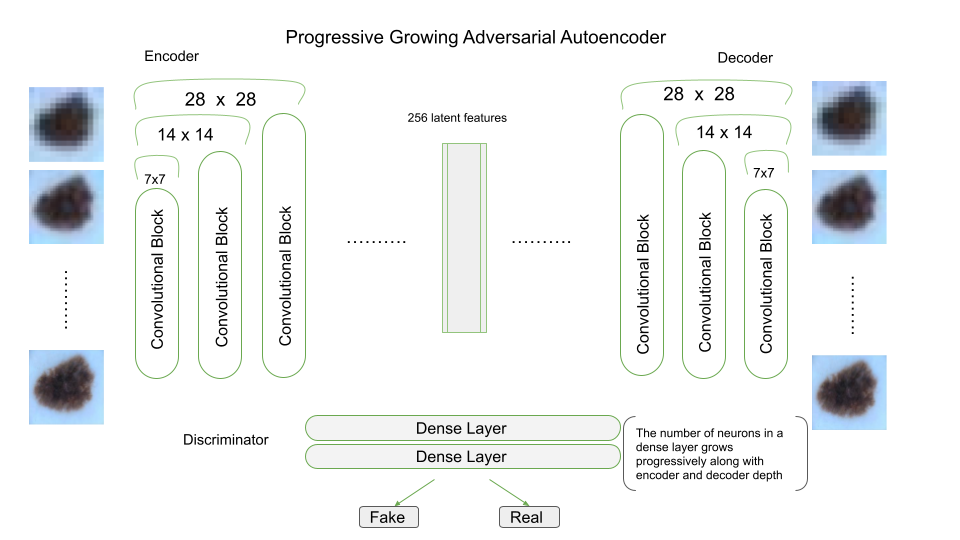}
    \caption{A Progressive Growing AAE network.}
    \label{fig:PGAAE}
\end{figure}

The PGAAE network paradigma is reported in Figure \ref{fig:PGAAE}.
First, we start by training a shallow AAE with just one convolutional block for both the encoder and decoder network. Such first AAE is trained to reconstruct skin lesion images resized to 7x7 pixels. Once the network is fully trained and optimized, it sends its weights to a second AAE with a deeper structure of two convolutional blocks for both the encoder and decoder network. This second AAE is trained with the same dataset resized to 14x14 pixels. At each step, weights are shared just over the common underlying structure and then kept trainable for future training steps, i.e. each AAE is initialized with the previous AAE weights and trained with images doubled in size.

After the sixth step, we have a fully trained AAE able to reproduce skin lesion images of size 224x224 pixels. In order to enhance the discriminator ability to discriminate between more and more complex images, at each step the discriminator grows in width: it consists of two dense layers with a progressive growing number of neurons (from 500 to 3000 with a step increase of 500 neurons after each phase) and a Leaky ReLu activation with parameter $0.2$. All discriminators end with a single neuron dense layer with a sigmoid activation.

Each convolutional block includes two identical sets of three layers: a \texttt{conv2d} layer with stride 3, filters ranging from 16 to 128, followed by a batch normalization with parameter 0.95 and a ReLu activation. Depending on whether we consider the encoder or the decoder network, a \texttt{max pooling} or an \texttt{up sampling} layer is attached at the end of each block.

\subsubsection{Denoising Autoencoder}
Another major issue affecting the training of generative models is the tendency of learning the identity function: if the autoencoder has more nodes in the hidden layer than inputs, then it can just learn the data and the output simply equals the input. Hence, it does not perform any useful representation learning or dimensionality reduction.

Denoising autoencoders \cite{denoising} are a stochastic version of basic autoencoders. A Denoising autoencoder attempts to address the identity function issue by randomly corrupting input images that the autoencoder must then reconstruct.
By making the training process and the reconstruction phase more challenging, it is proved that denoising autoencoders mitigate the identity function issue and learn more robust representations.

Another way to gain more generalization, reduce the vanishing gradient problem and improve convergence is to add noise also to the discriminator inputs \cite{noise_injection}. Denoising autoencoder, adversarial generative training ad noise injection have been used separately to improve autoencoder performance. We augment the adversarial model with both a denoising feature applied to the generator and a noise injection in to the discriminator.
The denoising feature was particularly helpful in achieving good reconstruction performance for latent space with 256 latent features.
We opted for a Gaussian noise with standard deviation $\sigma = 0.1$ (we tried different value ranging from $0.05$ to $0.9$: it seems there is no significant difference in reconstruction accuracy in the range $\sigma \in [0.1, 0.3]$; accuracy starts to deteriorate quickly after $\sigma=0.3$). Despite we are not interested in low dimensional latent space (reconstructed images would be too blurry and not enough variegate), we found no significant advantage in injecting noise in such cases.

\subsubsection{Mini Batch Discrimination}
Mini Batch Discrimination \cite{minibatch} was originally introduced as a technique to mitigate collapse of the generator network. It is a discriminative technique for generative adversarial networks between whole minibatches of samples rather than between individual samples.

The idea is for the discriminator to consider an entire batch of data, instead of looking to a single input data. The mode collapse is then much easier to spot, since the discriminator understands that whenever all the samples in a batch are very close to each other, the data has to be rejected. This forces the generator to produce many good outputs in each data batch.
An $L_1$ penalization norm is concatenated with the original input and fed to the discriminator last but one layer. Such penalization quantifies the closeness between data in the same minibatch, causing the discriminator to reject batches that are internally too similar. This discriminative technique along with the progressive growing network structure, helped to avoid the mode collapse for batches of small and middle size (16-64) at the cost of a small increase of discriminator parameters. Indeed, training AAE with small batches size increases the chances of falling in to a mode collapse.
However, small batch sizes are forced by hardware limitation due to high resolution images and high dimensional latent space.

Following \cite{minibatch}, a minibatch discrimination layer needs two hyperparameters to be fine tuned, namely $B$ and $C$ in the original paper, i.e. the number of discrimination kernels to use and the dimensionality of the space where closeness of samples is calculated. Hypothetically, the larger $B$ and $C$ are, the better results are obtained at the price of a lower computation speed. A good compromise between accuracy and speed was the choice $B=16$ and $C=5$.

\subsubsection{Performance}
After a thorough fine tuning of all three networks structures (encoder, decoder and discriminator) our PGAAE with 256 latent features achieves a reconstruction error measure through RMSE that ranges from $0.08$ to $0.24$ depending on whether we consider the most common or the most rare skin lesion class. Data augmentation was necessary to overcome scarcity and imbalance of the dataset. Mode collapse is greatly reduced, and we are able to generate variegate and good quality skin lesion images. ABELE explainer is now able to generate meaningful explanations.


\section{Explanations}
\label{sec:explanations}
The outcome of the classifier and the explanator are designed to present a compact visual interface to the user, organized into four panes: 1) the original image that was processed by the CNN and the label of the predicted classification; 2) an attention area that allows us to emphasize the areas that had a positive (brown color) or negative (green color) contribution to the classification; 3) a set of synthetic prototypes generated by the AAE that are classified with the same class of the input; 4) a counterexemplar, i.e. a synthetic image to present a prototype that is classified with a different class than the input.

\begin{figure}[t]
    \centering
    \includegraphics[width=\columnwidth]{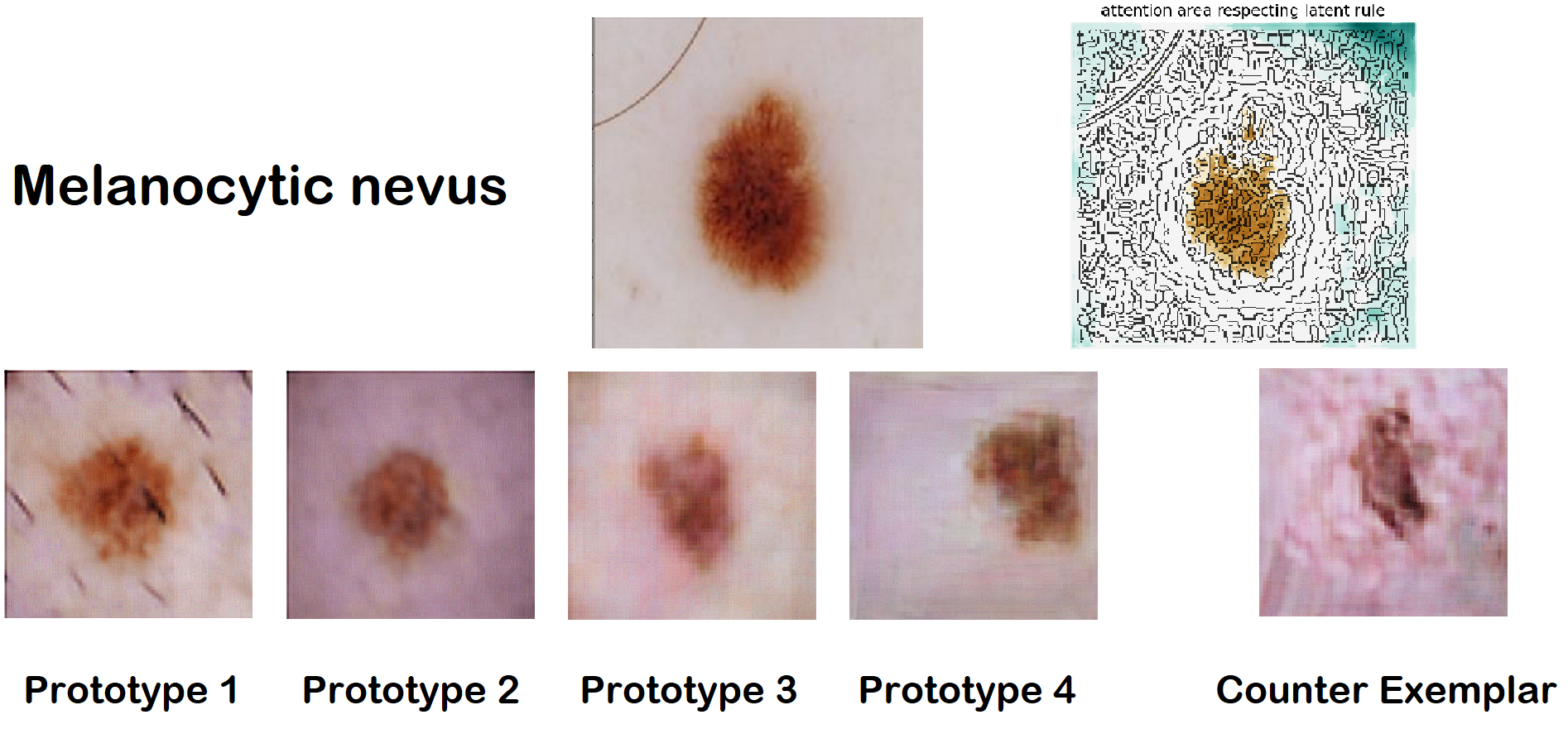}
    \caption{ABELE graphic explanation for a Melanocytic nevus}
    \label{fig:exp_sample_1}
\end{figure}

Figure \ref{fig:exp_sample_1} presents the outcome for an image, classified as \textit{Melanocytic nevus}. From the attention area the user is able to evaluate which parts of the input image where relevant for the CNN. The presentation of the results is complemented with the four prototypes: the four images are generated by the AAE, and they help the user to enforce the confidence with the decision yielded by the black-box, comparing the original image with the four exemplars. The counterexemplar, instead, has the function of probing the result of the black-box, by generating an image similar to the input but classified with a different class from the CNN.

ABELE generates statistics on the neighborhood of the input in the latent space. This information contributes to understand how the model space of the CNN is fragmented around the given input. This gives to the doctor a pulse of the classes that the black-box locates around the given instance. For the example in Figure~\ref{fig:exp_sample_1}, the statistics and rules on the latent space are the following:
\begin{equation*}
\begin{split}
    &\text{Neighborhood} \{NV:41 BCC:18 AK:4 BKL:26 DF:11\} \\
    &e = \{
        rules = \{ 7 > -1.01, 99 \leq 0.07, 225 > -0.75, 255 \leq -0.02, \\
        &238 > 0.15, 137 \leq -0.14 \} \rightarrow \{ class: NV \}\\
        &counter-rules = \{ \{ 7 \leq -1.01 \} \rightarrow \{ class: BCC \} \}
\}
\end{split}
\end{equation*}
Here the \textit{Neighborhood} entry shows the composition of the synthetic latent instances generated by the AAE. The rules and counter-rules are expressed in terms of ordinal positions of the dimensions in the latent space. Of course this representation is intended for internal use: it provides no accessible information to the human, but it may be exploited by the visual interface to implement an interactive refinement of the provided explanation





\section{Conclusion}
\label{sec:conclusion}
In this paper we propose a classification framework composed of a CNN model for the ISIC 2019 Challenge classification and a local-outcome explainer that produces exemplars and counter exemplars as an explanation. Our analytical framework contributes \textit{a)} to model a CNN to classify each input image as a class  among eight possibilities; \textit{b)} to create an explainer based on exemplars and counter exemplars synthesis that exploit an adversarial autoencoder (AAE) to produce the images for the explanation; \textit{c)} to tune the generation of exemplars to overcome the collapse mode issue to provide acceptable prototypes for the explanations. The CNN and the AAE follow distinct training processes, to demonstrate the possibility of using the explainer even with an external black-box, for example to test its fairness and dependability.

The analytical pipeline presented here is the core part of a wider system, where the interaction with the user should be further developed. In particular, we plan to enable an explorative process of the latent space of the input, by allowing the user to ask for additional exemplars or counter exemplars. Given the high impact on the cognitive layer of the user, we are designing a qualitative evaluation criteria with domain experts (i.e. doctors), to test the different dimensions of the explanations: usability, utility, comprehensibility, fidelity, etc.

\section*{Acknowledgment}
This work is partially supported by the European Community H2020 programme under the funding schemes:
G.A. 825619 \emph{AI4EU} ({\texttt{ai4eu.eu}}),
G.A. 952026 \emph{HumanE AI Net} ({\texttt{humane-ai.eu}}), 
G.A. 952215 \emph{TAILOR} ({\texttt{tailor-network.eu}}).

\bibliographystyle{IEEEtran} 
\bibliography{biblio}

\end{document}